\documentclass[12pt,a4paper,twocolumn]{article}
\usepackage{epsf,graphicx}
\newcommand{\goo}{\,\raisebox{-.5ex}{$\stackrel{>}{\scriptstyle\sim}$}\,}
\newcommand{\loo}{\,\raisebox{-.5ex}{$\stackrel{<}{\scriptstyle\sim}$}\,}

\makeatletter
\renewcommand{\section}{\@startsection{section}{1}{0mm}
{\baselineskip}%
{\baselineskip}{\normalfont\normalsize\bfseries\centering\uppercase}}%
\makeatother

\makeatletter
\renewcommand{\subsection}{\@startsection{subsection}{2}{0mm}
{\baselineskip}%
{\baselineskip}{\normalfont\normalsize\itshape}}%
\makeatother

\begin{document}
\twocolumn[
\begin{@twocolumnfalse}

\begin{centering} 
\large \textrm{\textbf{\uppercase{Statistical properties of fragment 
isospin in multifragmentation\\}}}
\end{centering}

\vspace*{0.5cm}

\begin{center} 
A.S.~Botvina$^{1,2}$
\end{center}

\vspace{0.3cm}
  $^{\;\;1}$ Gesellschaft f\"ur Schwerionenforschung mbH, D-64291 Darmstadt, 
Germany\\
  $^{\;\;2}$ Institute for Nuclear Research, Russian Academy of Sciences, 
117312 Moscow, Russia\\

\begin{abstract}
\footnotesize{
\noindent
It is shown that the chemical equilibrium condition of the system can 
be unambiguously identified by analyzing the isospin evolution of fragments 
produced in nuclear multifragmentation process. As far as the chemical 
equilibrium is established, the isotope production can be used for finding 
properties of nuclear matter and fragments in the freeze-out volume, 
for example, via the isoscaling phenomenon.}
\end{abstract}

\end{@twocolumnfalse}]

\section{Introduction}
Isotopic effects and chemical equilibrium in nuclear reactions are receiving 
increasing attention because they give a new insight into the nuclear 
matter properties at extreme conditions. This is of high current interest, 
in particular, for astrophysical applications. 
Reaction of multifragmentation of 
excited nuclei is very promising in this respect since it can 
be considered as manifestation of liquid-gas type phase transition in 
nuclei, therefore, it can directly address the nuclear matter behavior 
at subnuclear densities. A number of theoretical and experimental 
results was already reported on this subject [1--6]. 
Special interest to this problem arose after discovering the nuclear 
caloric curve based on the fragment isospin properties \cite{pochod95} 
and observation of similar isotopic temperatures in many experiments. 

\section{Statistical approach}
It is commonly accepted that the detailed balance principle is responsible 
for the chemical equilibrium between fragments of different sizes. 
Within statistical models, this can be written as a relationship between their 
chemical potentials. In the finite size systems 
we can directly implement this principle by taking into account all partitions 
with corresponding weights. For the case of the nuclear multifragmentation, 
both methods give very close results concerning the fragment isospin. 

The statistical multifragmentation model (SMM) is applied for the following 
analysis. This model has been developing since long ago and is successful in 
describing experimental data (see, e.g., references in 
\cite{botvina02,bond95}). Presently, many other statistical models 
are based on similar conceptions, therefore, the demonstrated physical 
results are representative for the whole statistical approach. 
The SMM assumes statistical equilibrium of highly excited nuclear systems 
at a low-density freeze-out stage $\rho \loo \rho_0/3$ 
($\rho_0\approx$0.15 fm$^{-3}$ is the normal nuclear density) \cite{bond95}. 
The SMM includes also the low energy de-excitation decay modes and 
predicts a natural transition to the multifragmentation regime 
from the compound nucleus decay \cite{botvina85}. 
All breakup channels (partitions) composed of nucleons and 
excited fragments are considered and the conservation of mass, charge, 
momentum and energy is taken into account. 
In the microcanonical 
treatment the statistical weight of decay channel j is given by 
$W_{\rm j} \propto exp~S_{\rm j}$, where $S_{\rm j}$ is the entropy of 
the system in channel j which is a function of the excitation 
energy $E_{\rm x}$, mass number $A_{\rm s}$, charge $Z_{\rm s}$ 
and other parameters of the source. 
Different breakup partitions are sampled according to their statistical 
weights uniformly in the phase space. After breakup, 
the fragments propagate independently in their mutual Coulomb field and 
undergo secondary decays. The deexcitation of the hot primary fragments 
proceeds via evaporation, fission, or Fermi-breakup \cite{botvina87}. 
At the freeze-out density the SMM treats hot fragments 
($A$, $Z$ are mass number and charge of the fragments) as 
nuclear matter pieces: 
Light fragments with $A\le 4$ are considered as stable
particles ("nuclear gas") with masses and spins taken from the nuclear tables. 
Fragments with $A > 4$ are treated as heated nuclear liquid  drops,
and their individual free energies $F_{AZ}$ 
are parameterized as a sum of the bulk,
surface, Coulomb and symmetry energy contributions
\begin{equation} \label{eq:faz}
F_{AZ}=F^{B}_{AZ}+F^{S}_{AZ}+E^{C}_{AZ}+E^{sym}_{AZ}.
\end{equation}
The standard expressions \cite{bond95} for these terms are: 
$F^{B}_{AZ}=(-W_0-T^2/\epsilon_0)A$, where the 
parameter $\epsilon_0$ is related to the level density, and 
$W_0$=16~MeV is the binding energy of nuclear matter; 
$F^{S}_{AZ}=B_0A^{2/3}(\frac{T^2_c-T^2}{T^2_c+T^2})^{5/4}$, where 
$B_0$=18~MeV is the surface coefficient, and $T_c$=18~MeV is the critical 
temperature of nuclear matter; $E^{C}_{AZ}=cZ^2/A^{1/3}$, where 
$c$ is the Coulomb parameter obtained in the Wigner-Seitz 
approximation, $c=(3/5)(e^2/r_0)(1-(\rho/\rho_0)^{1/3})$, with the charge 
unit $e$ and $r_0$=1.17 fm; $E^{sym}_{AZ}=\gamma (A-2Z)^2/A$, where 
$\gamma$=25~MeV is the symmetry energy parameter. These parameters 
are those of the Bethe-Weizs\"acker formula and correspond 
to the assumption of isolated fragments with normal density in the 
freeze-out volume, an assumption found to be quite successful in 
many applications. It is to be expected, however, that in a 
more realistic treatment primary fragments will have to be considered 
not only excited but also expanded and still 
subject to a residual nuclear interaction between them. 
These effects  can be accounted for in the fragment 
free energies by changing the corresponding liquid-drop parameters, 
provided such modifications are also indicated by the experimental data. 
For example, for the symmetry energy, 
this information may be obtained from the isoscaling phenomenon. 
An unresolved problem for all statistical models on 
the marked is the flow development observed at very high 
excitation energies. However, it seems not crucial for the isospin 
studies, since the flow influences mainly the kinetic energies of fragments 
and, possibly, their charge distributions. As far as the charge 
distributions are described, the isotope composition of the fragments 
can be reliably addressed within the statistical approach. 

\section{Evolution of the fragment isospin}
The crucial questions to be answered are: 1) what is 
the isospin distribution for fragments with fixed $A$ (or $Z$), 2) how 
does the isospin of fragments change with the isospin of the thermal source, 
3) how does the fragment isospin change with excitation energy of the source, 
when the fragment charge distribution evolves. 
In the grand canonical approximation \cite{botvina85}, the mean 
multiplicity of fragments is given by 
\begin{equation} \label{eq:naz}
\langle N_{AZ}\rangle
=g_{AZ}\frac{V_{f}}{\lambda_{T}^{3}}A^{\frac{3}{2}}\exp
\left[-\frac{F_{AZ}-\mu A-\nu Z}{T}\right],
\end{equation}
where $g_{AZ}$ is the degeneracy factor, 
$\lambda_{T}$ is the nucleon thermal wavelength, $V_f$ is the "free" volume, 
and $\mu$ and $\nu$ are the chemical potentials responsible for the 
mass and charge conservation in the system, respectively 
\cite{bond95,botvina85}. 
As shown in Ref.~\cite{botvina87}, 
the average charge $\langle Z_{A} \rangle$ of fragments with mass $A$
and the width $\sigma^{A}_{Z}$ of their charge distribution 
can be written as 
\begin{equation} \label{eq:za}
\langle Z_{A} \rangle  \simeq \frac{(4\gamma+\nu)A}{8\gamma +2cA^{2/3}},~~
\sigma^{A}_{Z} \approx \sqrt \frac{AT}{8\gamma}.
\end{equation}
The symmetry energy, Coulomb interaction and the chemical potential $\nu$ 
are directly responsible for the isospin of the produced fragments. 
As seen from Eq.~(\ref{eq:za}), the neutron content of the fragments 
increases with their mass number, because of the Coulomb term. This trend 
exists in the stable nuclei 
and it presents also in multifragmentation experimental data. 
\begin{figure}[ht]
\vspace{-0.2cm}
\hspace{-0.6cm}
\includegraphics[scale=0.53]{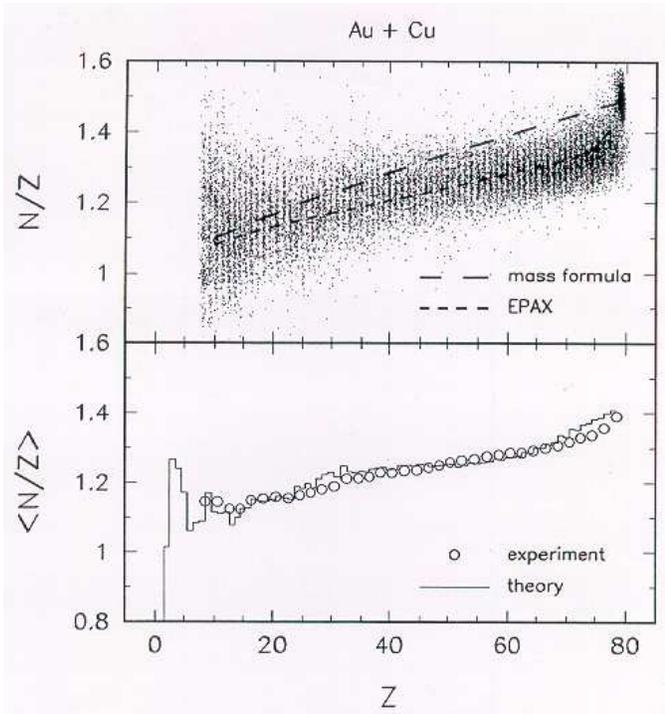}
\vspace{-0.4cm}
\caption{\small \label{fig:fig1}
Measured $N/Z$ (top) and mean value $\langle N/Z \rangle$ (bottom) as a 
function of the charge $Z$ of projectile fragments from the reaction 
$^{197}$Au~+~Cu at $E/A$=600~MeV. The lines in the top panel represent 
the valley of stability according to the Weizs\"acker mass formula 
(long dashed) 
and EPAX parameterization \cite{EPAX} (short dashed). In the bottom panel, 
the experimental data are given by the open circles and the prediction of 
the SMM by the histogram.
}
\end{figure}
One can see it from Fig.~1, where the neutron to proton ($N/Z$) ratio 
of fragments observed by ALADIN \cite{botvina95} at multifragmentation 
of Au-like sources is shown. The agreement of the SMM 
calculations with the data is obtained simultaneously with a very good 
description of the fragment partitions \cite{botvina95}, this is an 
important requirement for this kind of analysis.  

An intuitive expectation that the fragments produced from a neutron rich 
source are also neutron rich was confirmed in many experiments (see, e.g., 
\cite{milaz00,tsang01}). It is fully consistent with the statistical 
predictions: As calculations show \cite{botvina01,botvina02} the chemical 
potential $\nu$ decreases with the neutron excess of the source, therefore, 
this effect for intermediate mass fragments (IMF, charges Z=3--20) 
is explained by Eq.~(\ref{eq:za}). 

In Fig.~2 the $N/Z$ ratio of the fragments together with their mass 
distributions are shown. 
The calculations were done with the microcanonical Markov-chain SMM version 
\cite{botvina01} to take into account the mass and charge conservation 
explicitly. The results are consistent with the grand canonical 
ones (see also \cite{botvina02,bond95}), however, one can see that 
the finite-size effects favor the saturation of the $N/Z$ ratio of 
very large fragments with $A>A_{\rm s}/2$. This deviation from 
Eq.~(\ref{eq:za}) is important for low multiplicity 
channels at small $E_{\rm x}$. 

\begin{figure}
\vspace{-0.5cm}
\includegraphics[width=9cm]{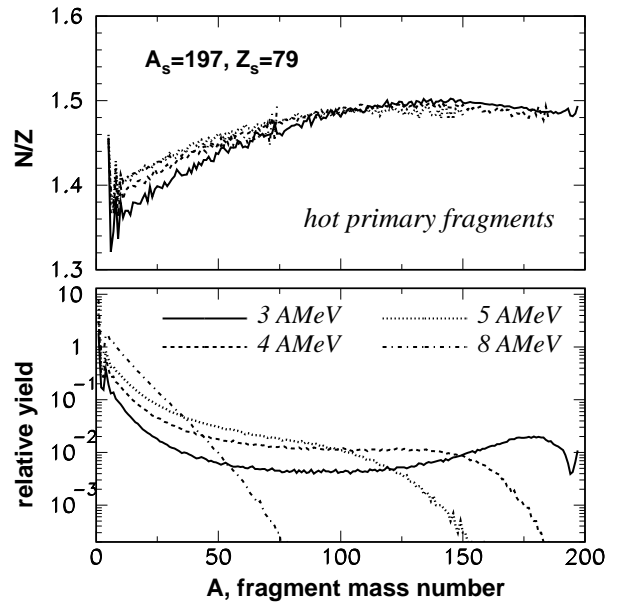}
\vspace{-0.5cm}
\caption{\small The neutron-to-proton ratio N/Z (top) and 
relative yield (bottom) of
hot primary fragments produced after the break-up of Au nuclei at
different excitation energies: 3 (solid lines), 4 (dashed lines),
5 (dotted lines) and 8 (dot-dashed lines) MeV per nucleon.}
\end{figure}

 The SMM predicts an interesting behavior of fragment isospin with changing 
fragment charge distribution. In Fig.~2 one can also see 
the evolution of the $N/Z$ ratio and mass distribution of 
fragments in the excitation energy range $E_{\rm x}$=3--8 MeV/nucleon. 
It is seen that the fragment mass distribution evolves from the U--shape, 
at the multifragmentation threshold $E_{\rm x}$=$E_{th} \approx 3$ 
MeV/nucleon, to an exponential fall-off at high energies. 
During this evolution the temperature reaches a "plateau" and is nearly 
constant \cite{pochod95,bond95}. 
As the energy increases the N/Z ratio of primary 
IMFs' increases, too. 
The reason is that the heaviest neutron-rich fragments are destroyed at 
increasing excitation energy, and most of their neutrons are bound in IMFs, 
since the number of free neutrons is still small at this stage. 
Generally, the $N/Z$ ratios of fragments decrease after 
their secondary deexcitation (e.g., compare Fig.~1 and Fig.~2). If the 
temperature increases considerably with the excitation energy 
(in the energy region of small $E_{\rm x} < E_{th}$) 
the $N/Z$ ratios of final cold fragments can decrease 
with the energy. However, when the temperature reaches the "plateau" 
the internal excitation energy of hot fragments does not change 
(see Eq.~(\ref{eq:faz})) 
and the final isotope composition becomes proportional to the primary fragment 
isospin. In this case, the effect of increasing the $N/Z$ ratio 
of IMFs can survive after the secondary deexcitation. 
The experimental data \cite{milaz00} show that this effect exists 
and is consistent with the SMM predictions. 
Therefore, it is a good tool for probing the freeze-out 
conditions and verification of the chemical equilibrium. 

It is important that the demonstrated isospin evolution takes place in 
the energy range 
which is usually associated with a liquid-gas type phase transition in 
finite nuclei \cite{bond95}. The realistic statistical model, 
designed to reproduce the experiment, does not 
support conclusions of some mean-field approaches 
(see Ref.~\cite{muller}) 
about separation of the nuclear matter into a neutron-rich gas and 
isospin symmetric "liquid" matter during the phase transition. 
On the contrary, the chemical equilibrium in finite nuclear systems 
suggests few free neutrons and neutron-rich big fragments 
at the freeze-out stage. That is consistent with experimental 
observations. 

\nobreak
\section{Isoscaling}

By using the chemical equilibrium conception as a guideline one can 
extract more information about fragments and nuclear matter properties 
from the analysis of experimental data. 
The scaling properties of cross sections for fragment production 
with respect to the isotopic composition of the emitting systems 
were studied long ago in light-ion induced reactions \cite{lozh92}. 
Recently, the isoscaling was also reported for heavy-ion reactions 
\cite{tsang01}. It is constituted by the exponential dependence of 
the production ratios $R_{21}$ for fragments with neutron number 
$N$ and charge $Z$ in reactions with different isospin asymmetry: 
\begin{equation}
R_{21} = \frac{Y_2(N,Z)}{Y_1(N,Z)} = C \cdot exp(N\cdot \alpha + Z\cdot
\beta), 
\label{eq:scalab}
\end{equation}
with three parameters C, $\alpha$ and $\beta$. 
Here $Y_2$ and $Y_1$ denote the yields from the more neutron rich 
and the more neutron poor reaction system, respectively. 
In some reactions, the parameters $\alpha$ and $\beta$ 
have the tendency to be 
quite similar in absolute magnitude but of opposite sign \cite{tsang01}. 
This suggests an approximate 
scaling with the third component of the isospin $t_3$=$(N-Z)/2$: 
\begin{equation}
R_{12} = \frac{Y_1(N,Z)}{Y_2(N,Z)} = C \cdot exp(-t_3 \cdot \beta_{t3}). 
\label{eq:scalboga}
\end{equation}
\begin{figure}[ht]
\vspace{-0.1cm}
\hspace{0.3cm}
\includegraphics[height=9cm]{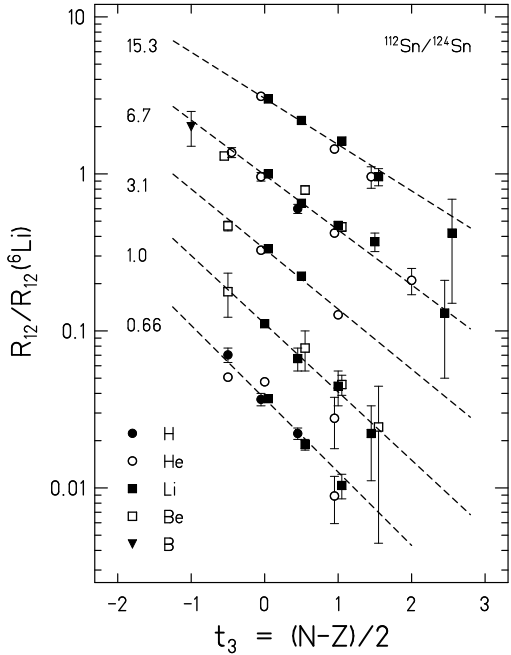}
\vspace{-0.3cm}
\caption{\small Normalized isotopic effect $R_{12}$
for five reactions, labeled with the projectile
energy (in GeV) and shifted by factors of three for clarity.
The symbols are for H, He, Li, Be, and B fragments.
The lines are the
results of the exponential fits according to Eq.~(\protect\ref{eq:scalboga}).}
\end{figure}
As an example, 
the isotope ratios measured in five pairs of reactions induced by light 
particles 
from protons of 0.66 GeV to $\alpha$-particles of 15.3 GeV incident energy on 
$^{112}$Sn and $^{124}$Sn targets, and normalized with respect to the ratio 
for $^{6}$Li, are shown in Fig.~3 (see Ref.~\cite{botvina02}). 
The parameter $\beta_{t3}$ decreases 
from 1.08 to 0.68 with increasing energy. 
The inverse scaling parameter 1/$\beta_{t3}$ is found to increase 
approximately in proportion to the isotope temperature deduced from the 
yields of He and Li isotopes. 

The isoscaling phenomenon arises naturally in a statistical fragmentation 
mechanism \cite{botvina02,tsang01a}. 
As evident from Eq.~(\ref{eq:naz}), for two systems 1 and 2 with different 
total mass numbers ($A_{1}$ and $A_{2}$) and charges ($Z_{1}$ and $Z_{2}$) 
but with the same temperature and density, 
the ratio of fragment yields produced in these systems is given 
by Eq.~(\ref{eq:scalab}) with parameters 
$\alpha=(\mu_1-\mu_2)/T$ and $\beta=((\mu_1-\mu_2)+(\nu_1-\nu_2))/T$. 
In the SMM the isoscaling parameters reflect properties of 
the produced fragments, in particular, the widths of their charge (mass) 
distributions (see Eq.~(\ref{eq:za})). As shown in Ref.~\cite{botvina02} 
the potential differences
depend essentially only on the coefficient $\gamma$ of the symmetry term
and on the isotopic compositions of the sources but not on the temperature: 
\begin{eqnarray} \label{eq:dmunu}
\mu_1 - \mu_2 \approx -4\gamma
(\frac{Z_{1}^2}{A_{1}^2}-\frac{Z_{2}^2}{A_{2}^2}),\nonumber\\
\nu_1 - \nu_2 \approx 8\gamma
(\frac{Z_{1}}{A_{1}}-\frac{Z_{2}}{A_{2}}).
\end{eqnarray}
This formula is a very good approximation (within 3\%) in the 
grand-canonics, and it was confirmed for the multifragmentation region 
($T \goo$5~MeV) with the microcanonical Markov chain SMM calculations. 
Moreover, since the temperature can be found in an independent way, 
e.g. as the isotope temperature, 
Eq.~(\ref{eq:dmunu}) provides an effective way for determination of the 
symmetry energy coefficient $\gamma$ via experimental $\alpha$ and 
$\beta$ parameters. 
\begin{figure}[ht]
\vspace{-0.15cm}
\includegraphics[width=9cm]{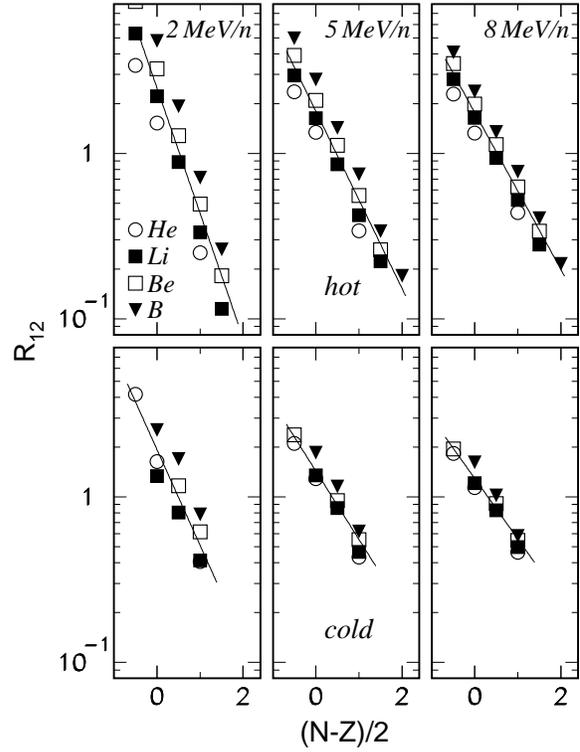}
\vspace{-0.7cm}
\caption{\small 
Ratios of isotope yields produced at the breakup of $^{112}$Sn and 
$^{124}$Sn sources from Markov-chain SMM calculations for three 
excitation energies $E_{\rm x}/A$~=~2, 5 
and 8~MeV. 
The top and bottom panels are for hot and cold fragments, respectively. 
The solid lines give the logarithmic slope parameters 
$\beta_{t3}$=1.74, 1.23, 1.09 for hot fragments, and 1.33, 0.95, 0.84 
for cold fragments. 
}
\end{figure}

A typical isoscaling obtained with the SMM is shown in Fig.~4. These 
calculations can be used to estimate quantitatively the influence of the 
secondary deexcitation on the observable parameters. 
The isoscaling parameters can change, 
and the corresponding correction to the extracted symmetry energy 
coefficient has to be applied. The full analysis was performed 
in Ref.~\cite{botvina02} and 
the value $\gamma \simeq 22.5$~MeV was reported, which has a tendency to 
be smaller than the conventional value. 
Provided it can be substantiated by
other data and analyses, this would indicate that the symmetry
part of the fragment binding energy is slightly weaker than that of
isolated nuclei. Fragments, as they are formed at breakup, may have a
lower than normal density.

\section{Conclusion}

Conception of chemical equilibrium is very fruitful for the theoretical 
description of multifragmentation. The chemical equilibrium can be 
unambiguously verified by analyzing isospin content of the fragments and 
its evolution together with the fragment yields. 
A convincing proof of this condition is an excellent 
description of the fragment charge partitions and the isotope 
distribution of fragments obtained in different experiments 
within statistical models (e.g., see \cite{bond95,botvina95}). 
A remarkable trend characterizing the chemical equilibrium is an increase 
of the $N/Z$ ratio with the mass number of fragments 
\cite{botvina01,botvina87}. This is quite instructive because dynamical 
processes can lead to the opposite ("decreasing") trend, as shown by 
M.~Colonna and M.Di~Toro in \cite{bali01}. 
Experimental observations of this trend may be obscured by the secondary 
deexcitation. However, by studying evolution of the $N/Z$ ratio of fragments 
with changing their charge distribution one can identify it for sure: 
The effect of increasing $N/Z$ of IMFs within the energy range of 
the phase transition, observed for the Au sources \cite{milaz00}, 
can be explained in the case of the "increasing" trend only. 
The chemical equilibrium conception can also 
provide a key for explanation of many other phenomena. For example, 
as shown in \cite{botvina01}, the midrapidity emission of neutron-rich IMF 
in peripheral heavy-ion collisions can be explained as a result of the 
Coulomb proximity of the sources and the angular momentum influence 
within the statistical picture of the two (projectile-like and 
target-like) decaying sources. 
An important advantage of this conception is that it allows for 
obtaining direct experimental information about nuclear properties 
at the break-up. A well-known example is the isotope temperatures, 
extracted from experimental data \cite{pochod95}. 
Another example is the isoscaling phenomenon, which is connected with the 
symmetry energy of nuclear matter. As shown here, the isoscaling gives 
a new way for probing the symmetry energy term of the fragments in the 
freeze-out volume. All these observations can be directly used in the 
astrophysical cases (supernovas, neutron stars), where the chemical 
equilibrium, is believed, to be established. 

{\it 
The author would like to thank W.~Trautmann, C.~Schwarz and J.~Lukasik 
for stimulating discussions and help in preparation of this presentation. 
}

\footnotesize{

}

\end{document}